\documentclass[letter]{aa} 
\usepackage{graphicx}
\usepackage{natbib}
\usepackage{lscape}
\usepackage{txfonts}
\begin{document}
   \title{High$-J$ CO emission in the Cepheus E protostellar outflow observed with SOFIA/GREAT}

   \author{A. I. G\'omez-Ruiz
          \inst{1}, 
          A. Gusdorf\inst{1,2}, S. Leurini\inst{1}, C. Codella\inst{3}, R. G\"usten\inst{1}, F. Wyrowski\inst{1}, M. A. Requena-Torres\inst{1}, C. Risacher\inst{1}, S. F. Wampfler\inst{4}
          }

   \institute{Max-Planck-Institut f\"ur Radioastronomie,
              Auf dem H\"ugel 69, 53121 Bonn, Germany
     \and
         LERMA, UMR 8112 du CNRS, Observatoire de Paris, \'Ecole Normale Sup\'erieure, 24 rue Lhomond, F75231 Paris Cedex 05, France
         \and
             INAF, Osservatorio Astrofisico di Arcetri, Largo E. Fermi 5, 50125 Firenze, Italy
        \and
            Institute for Astronomy, ETH Zurich, 8093, Zurich, Switzerland 
             }

   \date{Received ; accepted }

 
  \abstract
   {Owing to the high energy required for their excitation, high$-J$ CO transitions are a valuable tool for the study of protostellar jets and outflows. However, high spectral resolution observations of high$-J$ CO lines, which are essential to distinguish the different components in the line profiles, were impossible until the start of operations of the \emph{Herschel} space observatory and the Stratospheric Observatory for Infrared Astronomy (SOFIA).}
   {We present and analyze two spectrally resolved high$-J$ CO lines towards a protostellar outflow. We study the physical conditions, as a function of velocity, traced by such high-energy transitions in bipolar outflows.}
   {We selected the molecular outflow Cep E, driven by an intermediate-mass class 0 protostar. Using the GREAT receiver on board SOFIA, we observed the CO (12--11) and (13--12) transitions ($E_u \sim$ 430 and 500 K, respectively) towards one position in the blue lobe of this outflow, that had been known to display high-velocity molecular emission.}
   {We detect the outflow emission in both transitions, up to extremely high velocities ($\sim$ 100 km s$^{-1}$ with respect to the systemic velocity). We divide the line profiles into three velocity ranges that each have interesting spectral features: standard, intermediate, and extremely high-velocity. One distinct bullet is detected in each of the last two. A large velocity gradient analysis for these three velocity ranges provides constraints on the kinetic temperature and volume density of the emitting gas,  $\gtrsim$ 100 K and $\gtrsim$10$^4$ cm$^{-3}$, respectively. These results are in agreement with previous ISO observations and are comparable with results obtained by \emph{Herschel} for similar objects.}
   {High$-J$ CO lines are a good tracer of molecular bullets in protostellar outflows. Our analysis suggests that different physical conditions are at work in the intermediate velocity range compared with the standard and extremely high-velocity gas. }

\titlerunning{High$-J$ CO emission in Cep-E.}
\authorrunning{G\'omez-Ruiz, Gusdorf, Leurini, et al.}

   \maketitle
%

\section{Introduction} 

Observations over the past few decades have shown that during the star formation process, mass accretion is commonly associated with mass ejection in the form of jets. The interaction between the jet and the parent cloud generates shock fronts, that form large cavities called bipolar outflows. The simultaneous presence of jets and bipolar outflows, as well as accretion disks surrounding protostellar condensations, are commonly observed \citep{Arce07}, making them the basic elements of the star formation scenario. In this picture, different excitation regimes, related to the different structures, are revealed with a diversity of observational probes, such as continuum emission and molecular transitions at different wavelengths. In the jet/outflow system, at its temperature peak the shocked material is mainly radiatively cooled through the near-IR transitions of H$_2$, whereas the low-lying CO transitions trace the lower density, lower temperature ($<$ 50 K) post-shock gas. In-between those two extreme regimes and over the 100-2000~K range, the cooling of the gas occurs mainly through the emission of atomic and molecular lines such as [OI], H$_2$O, high$-J$ CO, OH, and H$_2$ transitions, which fall in the mid- to far-infrared spectral range \citep{Nisini99}. 

Traditionally, the physical conditions of the gas in protostellar outflows have been studied by means of ground-based observations of low-lying rotational transitions of the CO molecule (usually up to $J_{\rm u} = 3$, corresponding to E$_{\rm u} < 50$~K). In most cases, these low-energy transitions are more sensitive to the low excitation swept-up gas, not exclusively related with the jet and/or shock front. On the other hand, higher-lying transitions (E$_{\rm u} \gtrsim$100 K) probe the warm gas unambiguously related with the jet and/or hot spots in the shocked gas. However, transitions with relatively high$-J$ rotational quantum numbers (e.g. $J_{\rm u} \geq$ 9) fall into the window of the far-infrared regime that can not be observed from the ground. High$-J$ CO observations were pioneered with the \emph{Infrared Space Observatory} (ISO), which actually targeted several protostellar outflows \citep{Nisini99,Nisini00,Gia01}. These ISO observations revealed that the high$-J$ CO emission does indeed trace outflowing gas above 1000 K, but their poor spectral resolution did not allow us to probe either the presence of different spectral components, or the physical conditions of the gas as a function its velocity. In addition, their low spatial resolution ($\sim$ 80\arcsec~at wavelengths corresponding to its Long Wavelength Spectrometer, LWS) complicates the analysis owing to the presence of different structures within the same beam. Only recently, with \emph{Herschel} in operation and SOFIA starting operation, we had the opportunity to follow-up on these previous studies with high-resolution FIR spectroscopy and angular resolutions in the 15-20\arcsec~range. 

In this letter, we present one of the first studies of two high$-J$ CO spectrally resolved lines in a protostellar outflow, as an example of the potential of SOFIA observations with GREAT in the context of protostellar jets/outflows. For this study, we selected the protostellar outflow Cep E, in the Cepheus molecular cloud, since it is associated with a relatively nearby ($\sim$730~pc) intermediate-mass class~0 source \citep[IRAS 23011+6126:][]{Chini01} of 100 L$_{\odot}$, surrounded by an envelope structure of mass $\sim$ 18 M$_{\odot}$. The jet/outflow system itself has been intensively studied at multiple wavelengths \citep{Eis96,Lefloch96,Moro01,Smith03}. The strong CO wings and the evidence of jet-like components in the spectrum of Cep E outflow make it an ideal target for the study of the high$-J$ CO emission in jet/outflow systems.  

\section{Observations}

\begin{figure}[h!]
   \centering
   \includegraphics[width=8cm,angle=-90]{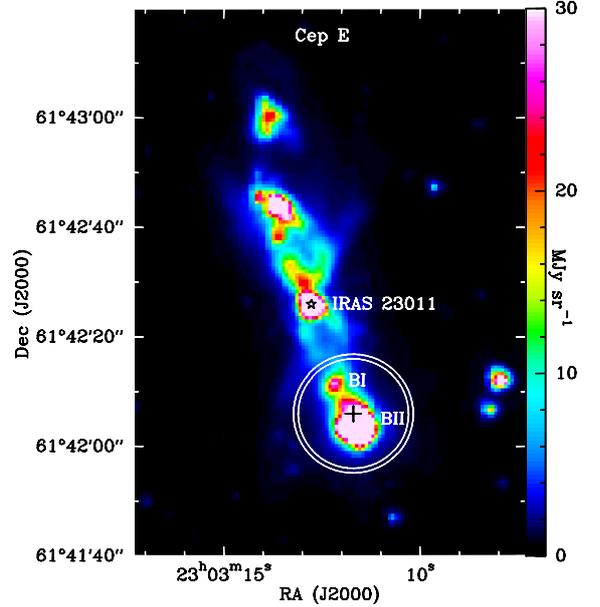}
      \caption{\emph{Spitzer}/IRAC band-two (4.5 $\mu$m) image of the Cep E protostellar outflow (retrieved from the \emph{Spitzer} archive). White circles mark the diffraction-limited SOFIA beams at the frequencies of CO (12--11) and (13-12) (decreasing beam size with increasing $J_{\rm u}$, see Table~\ref{trans}), centered at the brightest IR knot of the southern lobe, $\alpha =$23$^h$03$^m$11\fs70 $\delta =$$+$61$^{\circ}$42$'$06\farcs0 (black plus symbol). White letters label the two main knots inside our beams, BII and BI, as well as the Cep E central source (IRAS 23011: black star).
              }
         \label{irac-2}
   \end{figure}

The SOFIA telescope was pointed towards the southern lobe of the Cep E outflow (Fig.~\ref{irac-2}). As seen in Fig.~\ref{irac-2}, the SOFIA observations covered the infrared knots BI and BII. The observations were performed during basic science flights in July 2011, with the German Receiver for Astronomy at Terahertz Frequencies \citep[GREAT\footnote{GREAT is a development by the MPI f\"ur Radioastronomie and the KOSMA/Universit\"at zu K\"oln, in cooperation with the MPI f\"ur Sonnensystemforschung and the DLR Institut f\"ur Planetenforschung.}:][]{Heyminck12}. The L1 band of the GREAT instrument was used to observe, in two separate flights, the CO (12--11) and (13--12) transitions at $\sim$1.4 THz and $\sim$1.5 THz, in the lower and upper sidebands, respectively. In Table~\ref{trans}, the line frequencies, diffraction-limited beam sizes, time ON source ($t_{\rm ON}$), and atmospheric opacities ($\tau_{\rm atm}$) during the observations are presented. The front-end was connected to the AFFTS back-end \citep{Klein12}, providing a total bandwidth of 1.5 GHz and a spectral resolution of 212 kHz, that at the observed frequencies correspond to $\sim$ 300 km s$^{-1}$ and $\sim$0.4 km s$^{-1}$, respectively. Single pointing observations were done in double-beam chopped mode, with a chop throw of 60\arcsec~in RA (at 1 Hz). The pointing was established with the optical guide cameras, and was stable to 4\arcsec. The calibration uncertainty is within 20\%. From observations of Jupiter and Mars, a main-beam efficiency of 0.54$\pm$0.05 in the L1 band was determined \citep{Heyminck12}, that - with a forward efficiency of 0.95 - was used to convert antenna to main beam temperatures, $T_{\rm MB}$. Subsequent analysis of the data was done with CLASS within the GILDAS software\footnote{http://www.iram.fr/IRAMFR/GILDAS}, following the standard procedures of baseline subtraction and spectra averaging. A first-order baseline was subtracted from all spectra. 

Calibration was performed by carefully fitting the observed sky emission \citep{Guan12}. Unfortunately, both CO transitions suffer from residual terrestrial atmospheric features at SOFIA's flight altitude: the CO (12--11) detection bandwidth contains a strong water absorption (from the signal band) at positive velocities $>$60 km/s. This does not however affect the calibration of the pre-dominantly negative velocity emission towards the southern lobe. This is different for CO (13--12), whose calibration (and baselines) at extremely negative velocities (of lower than -~120~km s$^{-1}$) are more uncertain owing to an image band atmospheric feature. However, quite comparable integrated intensities for both lines at small S/N values supports our confidence in the adopted calibration procedure.

\begin{table}
\caption{The CO observations.}             
\label{trans}      
\begin{tabular}{l c c c c c c}     
\hline       
CO   & $\nu$$_0$   & $E_{\rm u}/k$   & HPBW & rms\tablefootmark{a} & $\tau_{\rm atm}$ & $t_{\rm ON}$\\
Line &  (GHz)      &    (K)      & (\arcsec) & (mK) & & (min)\\
\hline
(13--12)         &  1496.9922  & 503.1 & 20.0 & 82  & 0.03 & 15\\
(12--11)         &  1381.9951  & 431.3 & 21.7 & 96  & 0.16  & 7.4\\
(2--1)\tablefootmark{b}&  230.5380   & 16.6  & 20.0 & 32 & - & - \\
\hline            
\end{tabular}
\tablefoottext{a}{$T_{\rm{MB}}$ scale, at 3 km s$^{-1}$ spectral resolution.}
\tablefoottext{b}{Convolved to 20\arcsec~from original $\sim$ 11\arcsec~resolution (IRAM-30m), Lefloch et al. in prep.}
\end{table}

\section{Results}

\begin{figure}[h!]
   \centering
   \includegraphics[bb=50 80 539 717,width=8cm,height=8cm]{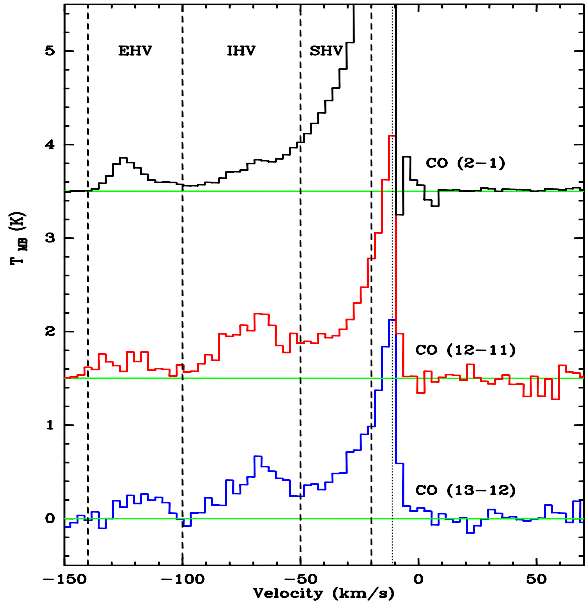}
      \caption{SOFIA/GREAT spectra of CO (12--11) and (13--12) in Cep E, together with the low$-J$ CO (2--1) transition, taken at the position shown in Fig.~\ref{irac-2}. The CO (13--12) and (2--1) spectra are taken from data with a spatial resolution of 20\arcsec, while the spatial resolution of the CO (12--11) is 21\farcs7. The two upper spectra have been shifted along the Y-axis. The green horizontal lines show the zero level of each spectrum. The dashed vertical lines indicate the limits of each velocity range defined in Table~\ref{tab:velint}. The dotted vertical line shows the systemic velocity ($-$11 km s$^{-1}$). The spectral resolution is 3 km s$^{-1}$ for all lines. 
              }
         \label{sofia-spec}
   \end{figure}

Together with the SOFIA/GREAT spectra, in Fig.~\ref{sofia-spec} we also show a CO (2--1) line profile taken at the corresponding position from a Nyquist sampled map, obtained with the IRAM-30m telescope (Lefloch et al., in prep). The CO (2--1) spectrum was convolved to a beam of 20\arcsec ~(i.e. the SOFIA beam at the (13--12) transition, see Table~\ref{trans}). From the CO spectra shown in Fig.~\ref{sofia-spec} at least three spectral features can be distinguished: two secondary peaks, at around $-125$ km~s$^{-1}$ and $-65$ km~s$^{-1}$, and a wing-like profile (smooth decrease of intensity towards high velocities) in the range from $-50$ km s$^{-1}$ to $-20$ km~s$^{-1}$. The secondary peaks are similar to the so-called molecular \lq bullets' often observed in the spectrum from class 0 protostars and likely related to the jet component \citep[e.g. L1448:][]{Bachiller90}. These \lq bullets' were reported previously in Cep E \citep{Lefloch96}. The wing-like feature is typical of the outflow phenomenon, mostly tracing the cavity walls crated by the jet, and usually referred to as the \lq standard' high-velocity component \citep[e.g.,][]{Bachiller90}. On the basis of these spectral features, in Fig.~\ref{sofia-spec} we divide all spectra into three velocity ranges: extremely high velocity (EHV: $-140$ to $-100$ km~s$^{-1}$), intermediate-to-high velocity (IHV: $-100$ to $-50$ km~s$^{-1}$), and standard high velocity (SHV: $-50$ to $-20$ km~s$^{-1}$). In Table~\ref{tab:velint}, we present the line intensities integrated within the defined velocity ranges. From the spectra in Fig.~\ref{sofia-spec}, one can see that the bullet-like profile in the IHV range is more prominent in high$-J$ than in low-$J$ CO, with the latter still being dominated by the wing-like profile that extends into this velocity range. On the other hand, the SHV component is stronger in low$-J$ than in high$-J$ CO. From the integrated intensities in Table~\ref{tab:velint}, we found that the (2--1)/(12--11) ratio is about 0.8, 0.4, and 2.2, for EHV, IHV, and SHV, respectively, indicating different excitation conditions in the different velocity ranges. These findings underline the necessity for velocity-resolved spectroscopy in the excited CO transitions, to identify the bullets as distinct features. The significant emission of the high$-J$ CO line at high-velocity indicates that it is very likely related to the jet component, unlike the wing component, which becomes weaker in these lines. Finally, the emission at low velocities is still considerable in our high$-J$ CO lines. This implies that the \emph{a priori} assumption made in the analysis of some ISO observations, namely that the excited CO emission originates exclusively from the high-velocity gas \citep[e.g.][]{Nisini00}, could be incorrect.

\begin{table}
\begin{minipage}{\columnwidth}
\caption{Velocity ranges and integrated intensities.}
\label{tab:velint} 
\renewcommand{\footnoterule}{}
\begin{tabular}{l c c c}     
\hline
CO     & \multicolumn{3}{c}{$\int T_{\rm MB}$dv (K km s$^{-1}$)\footnote{Statistical errors in parenthesis.}}\\
\cline{2-4}
Transition &    EHV    &  IHV  & SHV\\
\hline
13--12 & 4.8(0.9) & 14.9(1.0) & 16.1(0.8)\\
12--11 & 6.0(1.0) & 19.5(1.2) & 17.7(0.9)\\
2--1 & 5.1(0.2) & 8.8(0.3) & 38.9(0.2)\\  
\hline
\multicolumn{4}{l}{EHV: $-140$ to $-100$ km s$^{-1}$; IHV: $-100$ to $-50$ km s$^{-1}$}\\
\multicolumn{4}{l}{SHV: $-50$ to $-20$ km s$^{-1}$}\\
\hline
\end{tabular}
\end{minipage}
\end{table}  

\section{Discussion: physical conditions} 

   \begin{figure}
   \centering
   \includegraphics[width=6.2cm,angle=-90]{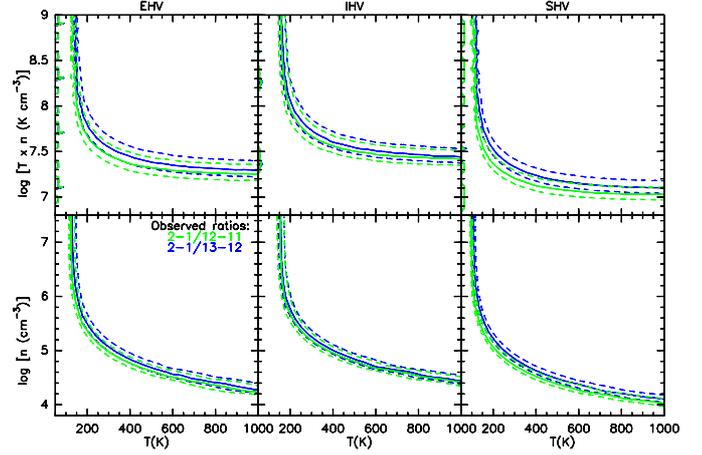}
      \caption{LVG results produced by the RADEX code for the observed intensity ratios in the EHV, IHV, and SHV ranges (left, middle, and right panels, respectively). The results are shown as $T$ vs. $n$ and $n \times T$ (thermal pressure) vs. $T$ plots (lower and upper panels, respectively). The CO (2--1)/(12--11) ratio is indicated in green and the CO (2--1)/(13--12) ratio in blue. The observed ratio is drawn with solid lines, while the line ratio uncertainty (including statistical and 20 \% of the calibration errors) with dashed lines. $N=2\times$10$^{17}$, 5$\times$10$^{17}$, and 7$\times$10$^{17}$ cm$^{-3}$; and $\Delta \varv$ = 40, 50, and 30 km~s$^{-1}$, for EHV, IHV, and SHV, respectively.
              }
         \label{lvg}
   \end{figure}

To constrain the physical conditions, we performed radiative transfer calculations with the RADEX code \citep{Tak07} based on the large velocity gradient (LVG) approximation and assuming a plane-parallel \lq slab' geometry for the escape probability formula. The molecular data were retrieved from the LAMDA data base\footnote{http://www.strw.leidenuniv.nl/$\sim$moldata/}. The collisional rate coefficients were adopted from \citet{Yang10}, who calculated the collisional rates between CO and H$_2$, incorporating energy levels up to $J = 40$ for kinetic temperatures of up to 3000 K. 

For the purposes of our study of the physical conditions as a function of velocity, we use the integrated intensities in the EHV, IHV, and SHV ranges (Table~\ref{tab:velint}). The value that we adopt for the linewidth, $\Delta \varv$, is directly inferred from the definition of our different velocity ranges (40, 50, and 30 km s$^{-1}$, for EHV, IHV and SHV, respectively). Our background radiation field is assumed to only be produced by the CMB. Then, our procedure consists in running, for each velocity range, a grid of RADEX models to compute the integrated intensities within a three-dimensional parameter space defined by $T$ (kinetic temperature), $n$ (H$_2$ volume density), and $N$(CO) (CO column density). 

The LVG results were analyzed based on the (2--1)/(12--11), (2--1)/(13--12) line ratios and the absolute integrated intensities of those three lines. We point out that owing to the use of these ratios the results are biased toward the lower excitations traced by the low$-J$ CO lines, and hence our results should be assumed to be lower limits. We use the IR size of the BI ($\sim$3\arcsec$\times$4\arcsec) and BII ($\sim$8\arcsec$\times$8\arcsec) knots to compute the brightness temperature, correcting for beam dilution effects. For the SHV component, we assumed the size of BII, while for the IHV and EHV components the size of BI was assumed. These assumptions are based on the knowledge that the EHV and IHV low$-J$ CO emission peaks around BI, while the SHV emission peaks around BII \citep{Hatchell99}. The LVG solutions were found when both the ratios and absolute integrated intensities match in the ($T$, $n$) plane. The $N$(CO) ranges constrained in this way are 4-9$\times$10$^{17}$, 4-6$\times$10$^{17}$, and 1-4$\times$10$^{17}$ cm$^{-2}$, for the SHV, IHV, and EHV components, respectively. We tested the influence of the size of the emission region on our $N$(CO) conclusions by varying the assumed size from a compact emission scenario, where the emission spreads over half the BI knot size, to a very extended case in which the filling factor is equal to one. Over this range, we found that $N$(CO) varies over up to two orders of magnitude. However, this change in $N$(CO) does not modify significantly the solution of the line ratios described below. In Fig.~\ref{lvg}, we show the RADEX results as plots of $T$ versus (vs.) $n$ and ($n \times T$) vs. $T$. In the latter, ($n \times T$) represents the thermal pressure, and is shown because in the LVG analysis the results are not as degenerate as for $T$ vs. $n$ (see Fig.~\ref{lvg}). The (2--1)/(12--11) and (2--1)/(13--12) ratios yield lower limits to the parameters of $T \gtrsim$ 100 K, $n \gtrsim$ 4.2$\times$10$^4$~cm$^{-3}$, and ($n \times T$) $\gtrsim 10^7$ K cm$^{-3}$. These lower limits indicate that the line ratios actually trace high densities and temperatures, but the remarkable result of our LVG analysis is that the thermal pressure tends to be higher in the IHV range than in both the EHV and SHV ranges. In addition, it is quiet a surprise that both EHV and SHV provide similar constraints, since it is usually found that the EHV component (probably related to the jet) has higher excitation than the SHV component (which is likely to trace the outflow cavity). 

The H$_2$O mapping observations at 183 GHz of \citet{Lefloch11}, which covered the blue lobe and the central region of Cep-E with an angular resolution of $\sim$13\arcsec, also showed prominent bullet-like spectral features in the IHV range. The strongest emission was found around the central region, while weaker and no emission at all was found at BI and BII, respectively. Together with additional SiO (8--7) observations, LVG calculations made for the gas in the central region yielded the physical conditions of $T\sim$ 200 K and $n\sim$10$^6$ cm$^{-3}$, which are comparable to the ones reported here, suggesting that the SiO, H$_2$O, and high$-J$ CO emission arises from gas with similar physical conditions, despite the different frequency range and positions probed. A correlation between the physical conditions of these molecules is an expected result given that the considered formation routes of SiO and H$_2$O associate themselves with warm and dense gas. However additional observations are required to verify the spatial correlation between these molecules and to provide more accurate determinations of the physical conditions.

Previous ISO observations of CO transitions from (14--13) up to (25--26), yielded kinetic temperatures and H$_2$ densities in the range 200-1200~K and 4$\times$10$^4$-4$\times$10$^6$~cm$^{-3}$, respectively. These constraints were obtained by fitting LVG models to the total integrated emission of these spectrally unresolved high$-J$ CO lines \citep{Gia01,Moro01}. It is remarkable that our lower limits are in good agreement with the values obtained from ISO, although our results are given for three distinct velocity ranges over the same global one. However, a more complete LVG analysis is needed to strengthen our conclusions with respect to previous ISO studies, and to more clearly probe the physical conditions prevailing in the observed regions. Such an analysis should be based on additional observations of both lower- and higher-$J$ CO transitions with  the GREAT spectrometer, such as the (11--10) and (16--15) ones, the latter would also allow us to make direct comparisons with ISO observations. 

Spectrally resolved \emph{Herschel}/HIFI observations have been used to study the emission in different velocity ranges in low-mass outflows by measuring the ratios of low$-$ to high$-J$ CO lines, which is an approach that is similar to our analysis. The kinetic temperatures generally inferred by those studies for the outflow components are in the range 100-200 K \citep{Yild10,Bjer11}. These values are similar to our lower limits. As already pointed out, the line ratios of a few high$-J$ CO lines with low$-J$ CO lines bias the results toward lower excitation conditions, and could in the end be inappropriate owing to the different physical component possibly related to each of them, as it has been proven in a few outflows \citep{Kempen10}. In this respect, additional observations of more high$-J$ CO lines, as well low- and mid$-J$ CO transitions, are required to test the possibility of the contribution of different physical components to the CO emission in outflows. Eventually, the combination of a maximum number of CO lines will provide the optimal way to shed light on the physical processes responsible for the existence of these components \citep[e.g.][]{Gusdorf12,Visser12}.


\section{Conclusions}

The principal conclusions of the present study are:

   \begin{enumerate}
      \item Our analysis of SOFIA/GREAT data has demonstrated that high$-J$ CO lines are a good tracer of molecular bullets in protostellar outflows.
        \item {The bullet at intermediate-to-high velocities has a higher level of excitation than the low and extremely high velocity gas, at the observed position of the Cep-E outflow.}
        \item The still considerable low-velocity emission in high$-J$ CO lines should be taken into account when modeling data.
        \item More and higher$-J$ CO transitions must be observed to break the degeneracy in the LVG solutions. 
   \end{enumerate}

\begin{acknowledgements}

Based on observations made with the NASA/DLR Stratospheric Observatory for Infrared Astronomy. SOFIA Science Mission Operations are conducted jointly by the Universities Space Research Association, Inc., under NASA contract NAS2-97001, and the Deutsches SOFIA Institut under DLR contract 50 OK 0901. We thank B. Lefloch for providing us with the CO data from IRAM-30m. A. Gusdorf acknowledges support by the grant ANR-09-BLAN-0231-01 from the French {\it Agence Nationale de la Recherche} as part of the SCHISM project.

\end{acknowledgements}

\bibliography{biblio}

\end{document}